\documentclass{desyproc}

\begin{document}
\title{\vspace{-3cm}
\hfill{\small{DESY 10-137}}\\[2cm]
Status of sub-GeV Hidden Particle Searches}

\author{{\slshape Sarah Andreas and Andreas Ringwald}\\[1ex]
Deutsches Elektronen-Synchrotron DESY, Notketra{\ss}e 85,
D-22607 Hamburg, Germany}

\contribID{andreas\_sarah}

\desyproc{DESY-PROC-2010-03}
\acronym{Patras 2010} 
\doi  

\maketitle

\begin{abstract}
Hidden sector particles with sub-GeV masses like hidden U(1)
gauge bosons, the NMSSM CP-odd Higgs, and other axion-like
particles are experimentally little constrained as they
interact only very weakly with the visible sector. For masses
below the muon threshold, we present constraints from meson
decays, $g - 2$ as well as beam-dump and reactor experiments.
The NMSSM CP-odd Higgs and generally any pseudoscalar is
required to be heavier than 210 MeV or couple to fermions much
weaker than the SM Higgs. Hidden photons are less constrained
and can be searched for at future fixed-target experiments,
e.g. HIPS~at~DESY.
\end{abstract}

\section{Motivation for a sub-GeV dark sector}

Hidden sectors are frequently proposed as part of the physics
beyond the standard model. Since their interactions with the
visible sectors are very weak, so are the current experimental
bounds. In fact, motivated both from a bottom-up and a top-down
perspective, those sectors might even contain light particles
with masses in the sub-GeV range that have so far escaped
detection. Among those weakly interacting slim particles
(WISPs) are hidden U(1) gauge bosons, the CP-odd Higgs of the
NMSSM, and other axion-like particles (ALPs).

Such particles are of great interest in many models that seek
to interpret recent terrestrial and astrophysical anomalies in
terms of dark matter (DM). The rise in the positron-fraction
with energy as observed by PAMELA (cf.~\cite{talk-Sparvoli})
and the deviation from the power-law in the $e^+ + e^-$
spectrum measured by FERMI (cf.~\cite{talk-Strigari}) together
with the absence of an excess in anti-protons require the DM
candidate to annihilate dominantly into leptons
\textit{(leptophil)} with a cross section much larger than the
one giving the correct relic abundance. Different direct
detection measurements like the annual modulation observed by
DAMA/LIBRA~\cite{talk-Cerulli} and the null results of CDMS and
XENON~\cite{talk-Balakishiyeva,talk-Oberlack} seem somewhat
contradicting. Consistency might still be possible if either
the DM candidate is light ($m \sim 5-10$ GeV) with elastic
scattering or heavy with excited states (mass splitting $\Delta
m \sim 100$ keV) generating inelastic scattering. Those
properties are challenging for standard DM candidates and
alternative scenarios like hidden sectors with light messenger
particles have been considered because of the following
advantageous features. A long range attractive force mediated
by such a light messenger generates a so called Sommerfeld
enhancement of the annihilation cross section. Dark matter
annihilation proceeding through this messenger -- if light
enough -- is naturally leptophilic due to kinematics. Inelastic
scattering on nuclei can also be mediated by such a light
particle. Possible examples of messenger particles that have
already been studied are ALPs like the NMSSM CP-odd
Higgs~\cite{Hooper:2009gm} and hidden U(1) photons of a generic
hidden
sector~\cite{ArkaniHamed:2008qn,Cheung:2009qd,Morrissey:2009ur,Cohen:2010kn}
or of asymmetric mirror worlds~\cite{An:2009vq}.

From a top-down perspective, hidden sectors appear naturally in
various supersymmetric models descending from string theory.
Mediator particles are generally weakly coupled to the visible
sector and can also be light. Specifically
in~\cite{Lebedev:2009ag} it was found that the heterotic string
can reproduce the NMSSM in a Peccei-Quinn limit with a light
Pseudo-Goldstone boson, an axion-like particle. The breaking of
larger groups down to the SM gauge group can in general yield
hidden U(1) symmetries which may remain unbroken down to small
energy scales. Their hidden photon may be light and couple
weakly to the visible sector through kinetic
mixing~\cite{Holdom:1985ag,Goodsell:2009xc}.

In the following we present various constraints on the NMSSM
CP-odd Higgs as representative of an axion-like particle and
the hidden photon for masses below the muon threshold.

\section{NMSSM CP-odd Higgs}

The extension of the MSSM with an additional scalar field $S$
to the NMSSM has been motivated as it solves the $\mu$-problem
by replacing the $\mu$-parameter with a SM singlet
$S$~\cite{Ellis:1988er}. Additionally, the enlargement of the
particle content by an additional CP-odd Higgs $A^0$ alleviates
the little hierarchy problem if $A^0$ is light by opening an
additional Higgs decay channel $h \to 2 A^0$, thereby reducing
the LEP limit on the Higgs mass. We focus our analysis on the
$Z_3$-symmetric NMSSM, a special version without direct
$\mu$-term, with superpotential
\begin{equation}
W = \lambda S H_u H_d + \frac{1}{3} \kappa S^3.   \nonumber
\end{equation}

In the limit $\kappa \rightarrow 0$, the Higgs potential
possesses an approximate Peccei-Quinn symmetry and a naturally
light pseudoscalar $A^0$ arises with $m_{A^0}^2 \simeq \kappa
\cdot \mathcal{O}(\mathrm{EW \, scale})^2$ where $\kappa \ll
1$. In the heterotic string example of~\cite{Lebedev:2009ag},
$\kappa$ can be as small as $10^{-6}$ resulting in a $100$~MeV
pseudoscalar. Its couplings to fermions are according
to~\cite{Dermisek:2010mg} given by
\begin{equation}
\Delta\mathcal{L} = - i \frac{g}{2 m_W} \ C_{Aff} \ \biggl(m_d~ \bar{d}
\gamma_5 d + \frac{1}{\tan^2 \beta} m_u~ \bar{u} \gamma_5 u + m_l~ \bar{l}
\gamma_5 l \biggr) \ A^0. \nonumber
\end{equation}
We treat $C_{Aff}$ as free parameter focusing on the range
$10^{-2} \lesssim C_{Aff} \lesssim 10^2$ to avoid violation of
perturbativity and/or finetuning and summarize the constraints
derived in~\cite{Andreas:2010ms} in the following.

\begin{wrapfigure}{r}{0.7\textwidth}
\centerline{
\includegraphics[width=0.342\textwidth]{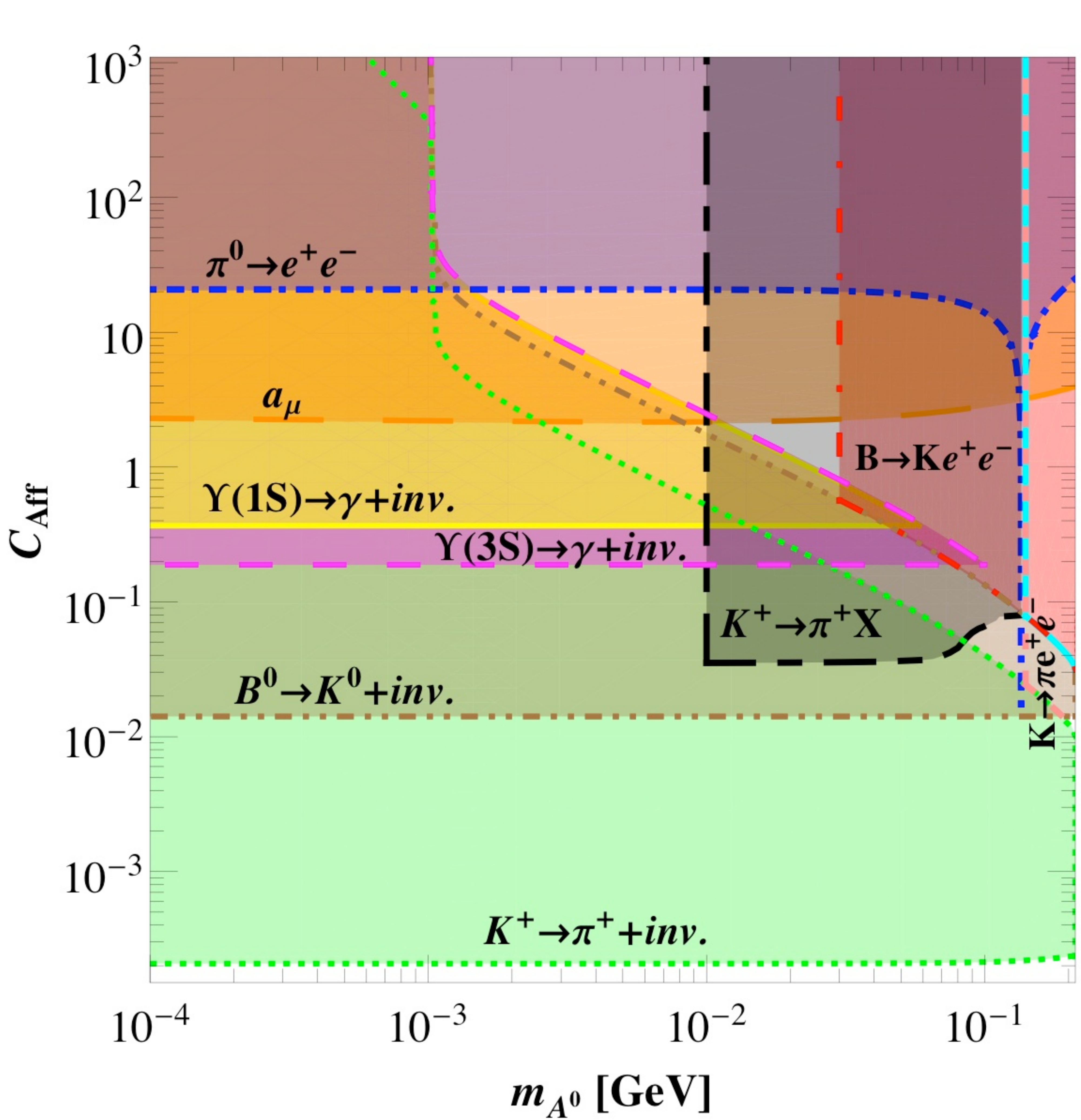}
\includegraphics[width=0.342\textwidth]{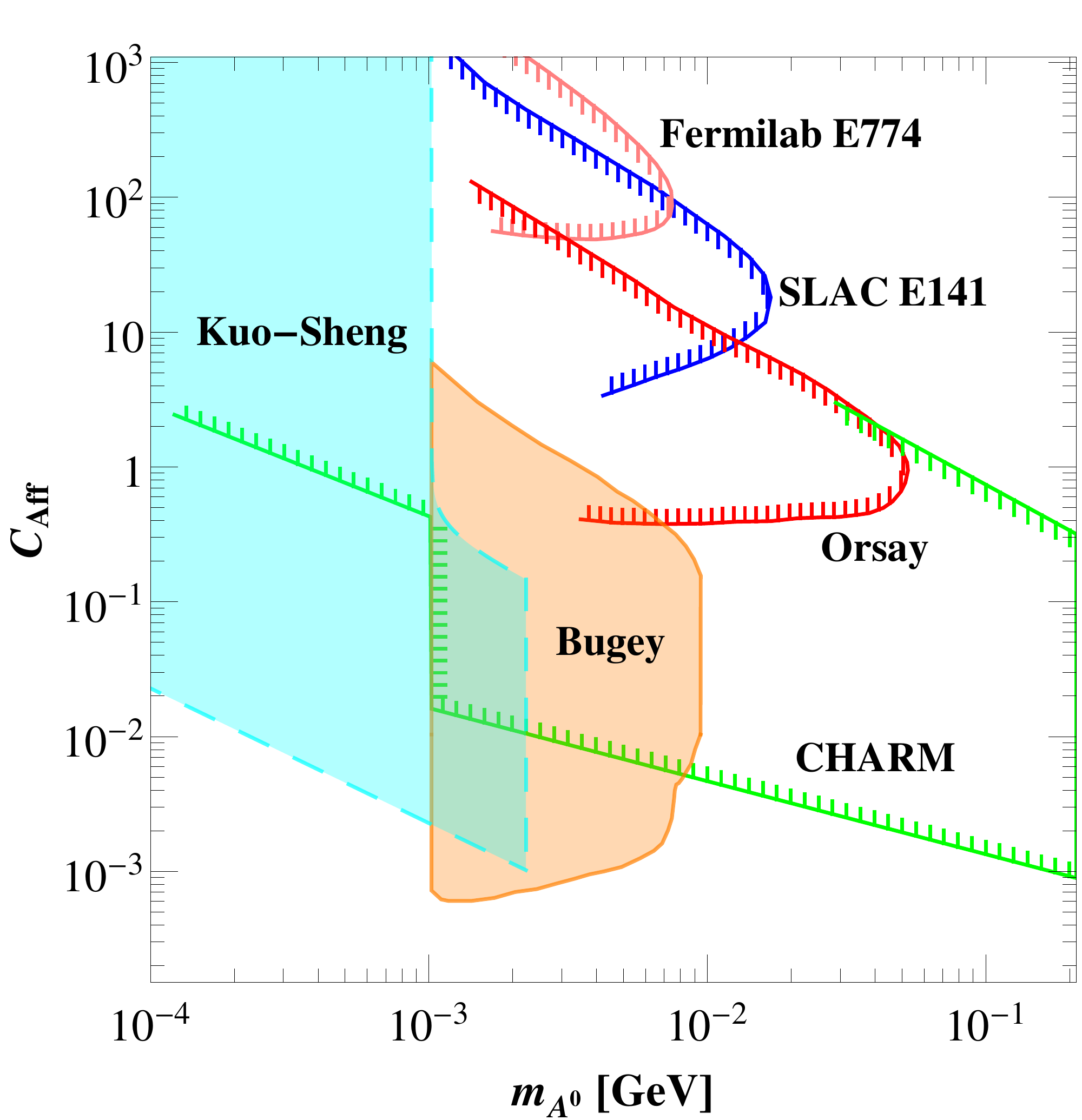}}
\caption{Excluded regions for the NMSSM CP-odd
Higgs~\cite{Andreas:2010ms}.} \label{Fig:NMSSM}
\vspace{-1cm}
\end{wrapfigure}

Different meson-decays set bounds for two distinct cases
depending on the lifetime of $A^0$. If it is sufficiently long
lived to escape the detector, invisible decays $X \! \to Y +
A^0 \to Y + $ inv. place limits requiring $\Gamma^{X \to Y A^0}
/ \Gamma^{\rm tot} < \mathcal{B}^{\rm exp}_{\rm inv}$. Larger
values of $C_{Aff}$ for which $A^0$ decays within the detector
are constrained by visible decays $X \! \to Y + A^0 \to Y + e^+
e^-$ demanding \mbox{$\rm{BR}^{X \to Y A^0} \rm{BR}^{A^0 \to
e^+ e^-} \!\!\!\! < \mathcal{B}^{\rm exp}_{e^+ e^-}$}. Together
with the limit from a search for a peak in the $\pi^+$~momentum
spectrum in $K^+ \! \to \pi^+ + X$, meson decays cover most of
the parameter space in Fig~\ref{Fig:NMSSM}.

Complementary constraints arising from the pion-decay $\pi^0
\to e^+ e^-$ and the muon anomalous magnetic moment $a_\mu$
completely close the available parameter space. The former
process which proceeds in the SM through loop diagrams receives
a tree level contribution from $A^0$ and sets a limit requiring
$\Gamma^{\pi^0 \overset{A^0}{\rightarrow} e^+ e^-} /
\Gamma^{\rm tot} \! < \mathcal{B}^{\rm exp}_{\pi^0 \to e^+
e^-}$. As there are several NMSSM contributions to $a_\mu$ of
both signs, even though the negative loop-contribution from
$A^0$ worsens the current discrepancy $a_\mu^{\rm exp} >
a_\mu^{\rm SM}$, we derive a constraint demanding $A^0$ not to
worsen it beyond $5\sigma$.

Additional constraints can be derived from beam-dump and
reactor experiments (lines and shaded regions, respectively, in
Fig.\ref{Fig:NMSSM}, right) searching for the decay $A^0 \to
e^+ e^-$. Like any ALP, $A^0$ can be emitted in the former via
bremsstrahlung from an $e$- or $p$-beam and in the latter in
place of photons in transitions between nuclear levels.

In summary, for masses below the muon threshold, the CP-odd
Higgs is excluded or required to couple to matter at least 4
orders of magnitude weaker than the SM Higgs which can hardly
be achieved in the NMSSM. Those constraints as they are plotted
in Fig.~\ref{Fig:NMSSM} apply in general to the coupling of a
light pseudoscalar to matter.

\vspace{0.3cm}

\section{Hidden U(1) gauge boson}

Many SM extensions contain additional U(1) symmetries in the
hidden sector under which the SM is neutral. The corresponding
gauge boson, the hidden photon $\gamma'$ and the ordinary
photon kinetically mix~\cite{Holdom:1985ag,talk-Redondo}
induced by loops of heavy particles charged under both U(1)
groups.

The most general Lagrangian is
\begin{equation}
{\cal L} = -\frac{1}{4}F_{\mu \nu}F^{\mu \nu} - \frac{1}{4}X_{\mu \nu}X^{\mu \nu} +
\frac{\chi}{2} X_{\mu \nu} F^{\mu \nu} + \frac{m_{\gamma^{\prime}}^2}{2}  X_{\mu}X^\mu \nonumber
\end{equation}
where $F^{\mu \nu}$ is the ususal electromagnetic field
strength and $X^{\mu \nu}$ the one corresponding to the hidden
gauge field $X^\mu$. The kinetic mixing $\chi$ is typically of
the size of a radiative correction $\sim \mathcal{O}(10^{-4} -
10^{-3})$. Kinetic mixing allows $\gamma'$ to couple and decay
to SM fermions thereby making it accessible for experimental
searches, the constraints of which are presented in the
following.

\begin{wrapfigure}{r}{0.7\textwidth}
\vspace{-0.4cm} \centerline{
\includegraphics[width=0.342\textwidth]{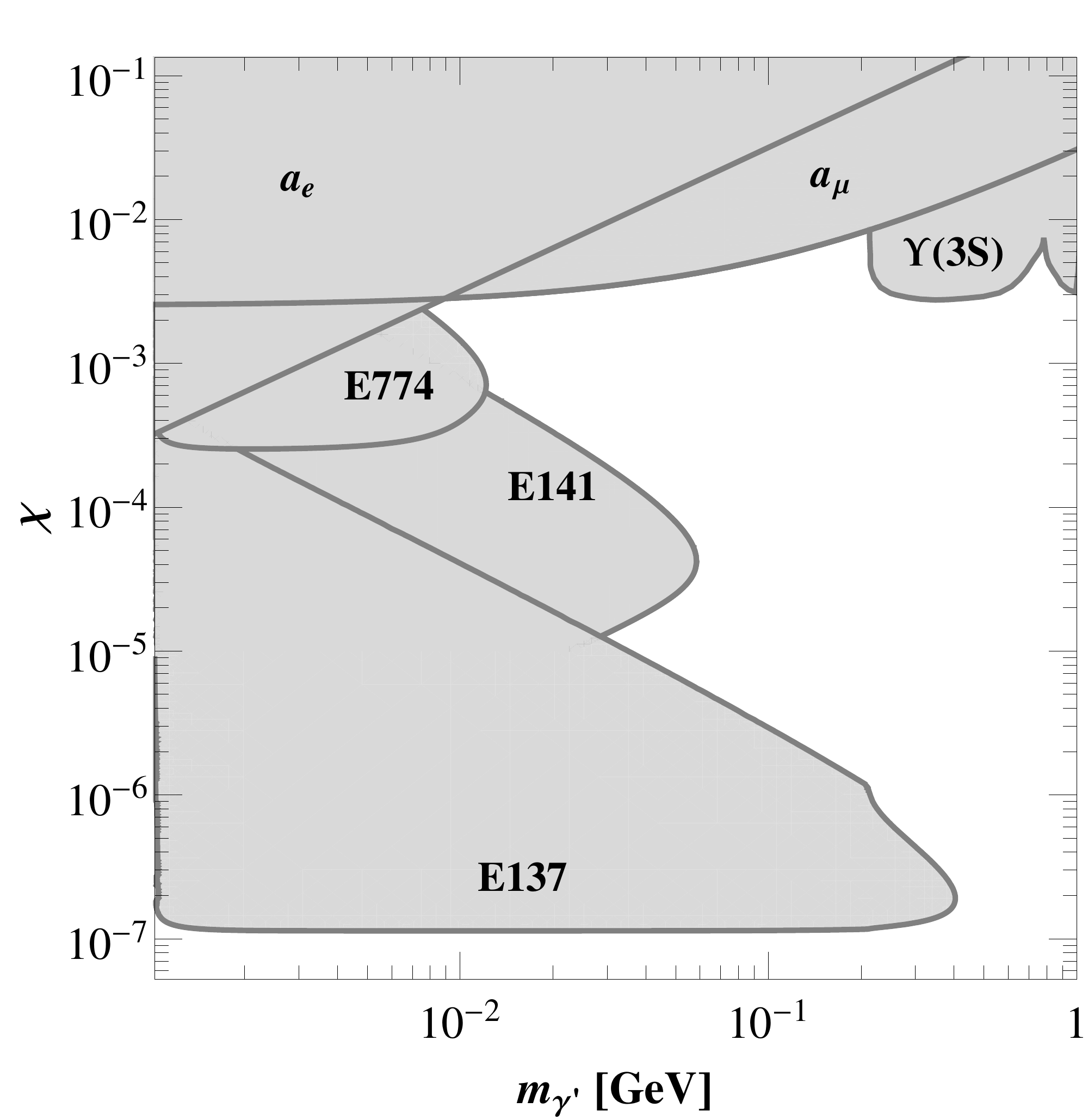}
\includegraphics[width=0.342\textwidth]{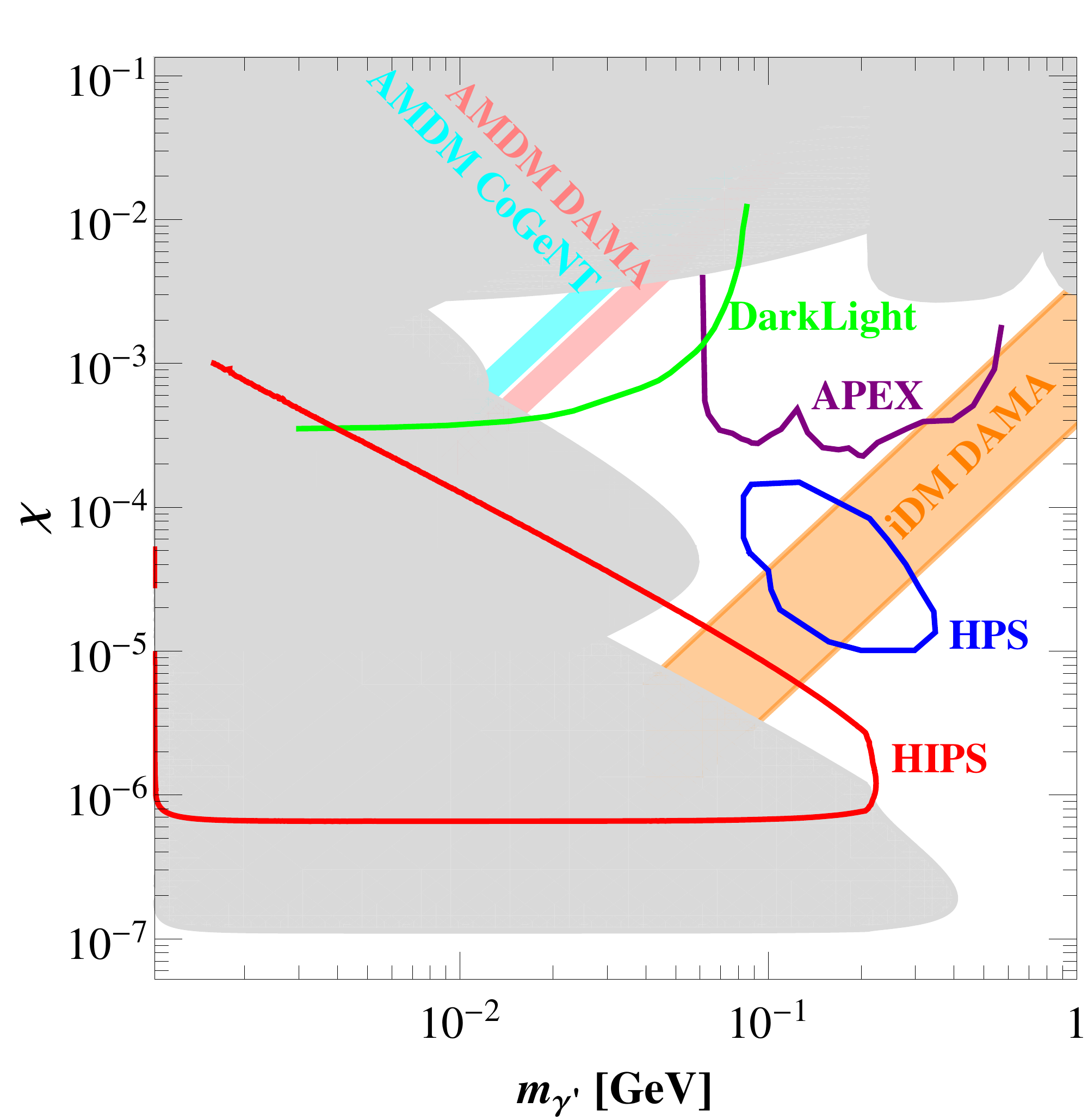}}
\caption{Exclusion regions \textit{(left)} as well as projected
sensitivities and phenomenological motivations \textit{(right)}
for the hidden photon.}\label{Fig:HP} \vspace{-0.4cm}
\end{wrapfigure}

Similarly to the CP-odd Higgs, limits arise from one-loop
contributions of the hidden photon to the muon and electron
anomalous magnetic moment~\cite{Pospelov:2008zw}. Also
beam-dump experiments in which $\gamma'$ is emitted through
bremsstrahlung \newline from an $e$-beam can set constraints by
searching for the decay $\gamma' \to e^+
e^-$~\cite{Bjorken:2009mm}. The resulting limits (shaded in
Fig.~\ref{Fig:HP}) leave an unexplored region in the parameter
space which is best explored by fixed-target
experiments~\cite{Bjorken:2009mm,Freytsis:2009bh}. Dedicated
proposals are being developed at DESY (HIPS, see
also~\cite{talk-Mnich}), JLab (APEX~\cite{talk-Afanasev},
HPS~\cite{talkSLAC-HPS},
DarkLight~\cite{Freytsis:2009bh,talkSLAC-DarkLight}), and MAMI,
with complementary sensitivities (cf. lines in
Fig.~\ref{Fig:HP}, right).

The whole allowed parameter range in Fig.~\ref{Fig:HP} is
phenomenologically interesting for DM with dark photons of a
generic hidden U(1)~\cite{ArkaniHamed:2008qn} or mirror photons
in asymmetric mirror DM models (AMDM)~\cite{An:2009vq}. The
former can reproduce DAMA for inelastic DM~\cite{Essig:2009nc}
(orange ``iDM'' band) and achieve naturally the leptophilic DM
annihilation required for PAMELA~\cite{Meade:2009iu}, while the
latter is able to explain the DAMA and CoGeNT measurements with
mirror neutrons as DM~\cite{An:2010kc} (colored ``ADMD''
bands).

\section{Conclusions}
Hidden sectors are well motivated by DM, SM extensions, and
string theory. They might contain light particles that despite
their very weak couplings to the SM can be constrained
experimentally. In particular, the NMSSM CP-odd Higgs has to be
heavier than 210 MeV or couple much weaker to fermions than the
SM Higgs. Hidden photons on the contrary are less constrained
and can be searched for in complementary experiments at DESY,
JLab and MAMI.


\begin{footnotesize}


\end{footnotesize}


\end{document}